\documentclass[aps,prd,twocolumn,showpacs,nofootinbib]{revtex4}

\usepackage[utf8]{inputenc}
\usepackage[T1]{fontenc}

\usepackage{latexsym}
\usepackage{amsmath,amsfonts}
\usepackage{amsbsy}
\usepackage{mathrsfs}
\usepackage{color}

\usepackage{psfrag}

\usepackage{enumerate}

\usepackage{amsmath,amssymb,calc,amsfonts}
\usepackage{latexsym}
\usepackage{hyperref}
\usepackage{ulem}

\usepackage{graphicx,calc,epsfig}
\def\ut#1{\rlap{\lower1ex\hbox{$\sim$}}#1{}}

\newcommand{\be}{\nopagebreak[3]\begin{equation}}
\newcommand{\ee}{\end{equation}}
\newcommand{\ba}{\nopagebreak[3]\begin{eqnarray}}
\newcommand{\ea}{\end{eqnarray}}

\DeclareFontFamily{U}{rsfs}{}         
\DeclareFontShape{U}{rsfs}{m}{n}{<5> rsfs5 <6><7> rsfs7          %
  <8><9><10><10.95><12><14.4><17.28><20.74><24.88> rsfs10}{}     %
\DeclareMathAlphabet{\mathfs}{U}{rsfs}{m}{n}                     %
\newcommand{\mfs}[1]{\mathfs {#1}}                               %
\newcommand{\n}{{\nonumber}}
\newcommand{\va}{\scriptscriptstyle}

\newcommand{\sI}{{\mfs I}}\newcommand{\sO}{{\mfs O}}

\def\i{i}

\def\pb#1{\rlap{\lower1.5ex\hbox{$\longleftarrow$}}{#1}}
\def\dpb#1{\rlap{\lower1.5ex\hbox{$\Longleftarrow$}}{#1}}
\def\spb#1{\rlap{\lower1.5ex\hbox{$\leftarrow$}}{#1}}
\def\sdpb#1{\rlap{\lower1.5ex\hbox{$\Leftarrow$}}{#1}}

\definecolor{blue}{rgb}{0,0,1}
\definecolor{green}{rgb}{0,0.6,0.5}
\definecolor{red}{rgb}{1,0,0}
\definecolor{vio}{rgb}{0.66,0,1}
\definecolor{ama}{rgb}{1,1,0}

\begin{document}

\title{
Spherically symmetric black holes and affine-null metric formulation of Einstein's equations 
}

\date{\today}

\author{Emanuel Gallo$^1$, Carlos Kozameh$^1$, Thomas M\"adler$^2$, Osvaldo M. Moreschi$^1$ and Alejandro Perez$^3$}

\affiliation{$^1$FaMAF, UNC; Instituto de F\'{i}sica Enrique Gaviola (IFEG), CONICET, \\
Ciudad Universitaria, (5000) C\'ordoba, Argentina. }

\affiliation{$^2$Escuela de Obras Civiles and N\'ucleo de Astronom\'ia, Facultad de Ingenier\'{i}a y Ciencias, Universidad Diego Portales, Avenida Ej\'{e}rcito
Libertador 441, Casilla 298-V, Santiago, Chile. }

\affiliation{$^3$ 
Aix Marseille Univ, Université de Toulon, CNRS, CPT, 13000 Marseille, France}

\begin{abstract}
The definition of well-behaved coordinate charts for black hole spacetimes can be tricky, as they can  lead for example to either unphysical  coordinate singularities in  the metric (e.g. $r=2M$ in the  Schwarzschild black hole)  or to an implicit dependence of the chosen coordinate to physical relevant coordinates (e.g. the dependence of the null coordinates in the Kruskal metric).
Here we discuss two approaches for coordinate choices in spherical symmetry allowing us to discuss explicitly "solitary"
and spherically symmetric black holes from a regular horizon to null infinity.
The first approach relies on a construction of a regular null coordinate (where regular is meant as being defined from the horizon to null infinity) given an explicit solution of the Einstein-matter equations. The second approach is based on an affine-null formulation of the Einstein equations
and 
the respective characteristic initial value problem. In particular, we present a
derivation of 
the Reissner-Nordstr\"{o}m black holes 
expressed in terms of these regular coordinates.
\end{abstract}


\maketitle

\section{Introduction}\label{sec:intro}
The classical theory of General Relativity (GR) predicts the existence of fascinating compact objects like black holes. 
{They} are, roughly speaking, regions of spacetime in which our of the physical laws breaks down and from which no information can escape. That black holes are not just an academic mathematical solution of Einstein's field equations find one of its recognition in the 2020 Nobel prize\footnote{Not to mention the 2017 Nobel prize given to The Ligo collaboration as well as direct measurement of a black hole shadow  of the supermassive black hole in the galaxy M87 by the Event horizon telescope.} award being given to the two astrophysisicts, Andrea Ghez and Rainer Genzel, and the mathematical physicist Rodger Penrose. While the astrophysicists received the award for the (indirect ) astronomical observation of the central black hole in the Milky Way, Penrose received it for ``for the discovery that black hole formation is a robust prediction of the general theory of relativity. \cite{Nobel}''. But Penrose contributions to our understanding of black holes go further than this, as his research  provided most of mathematical tools we use nowadays to analyze black hole spacetimes. One of the defining properties of a black hole is the presence of an event horizon, a null hypersurface separating the interior of a black hole from an  external observer.

Since the analysis of a spacetime involves the definition of a spacetime chart, adapted coordinates may be  given in a way that the metric $g_{ab}$ is either well defined or singular if it is evaluated at the horizon. On one hand, the classical  example in  a spherically  symmetric spacetime for singular coordinates at the event horizon are the Schwarzschild coordinates or the Eddington-Finkelstein coordinates. In the first case, the metric components blows up at the event horizon located at the radius $r=2M$, where $M$ is the mass of the black hole; while in the second case, the null coordinate diverges at the horizon. On the other hand, one example of well defined coordinates in spherical symmetry at the black hole horizon is given by the Kruskal-Szekeres coordinates, which are globally well defined and only singular at the central singularity $r=0$.
If we are interested in studying matter fields in the vicinity of the horizon, we can see that it is of importance to have well-defined coordinates at the horizon as otherwise no proper statements on the physical behaviour of those fields can be made. However, having well behaved  coordinates at the horizon is one  side of the story, only, because we do not only want to study fields at and near the horizon. We also want to know how these fields behave far away from it, as this is the region where the external observer is making his/her measurements of the dynamical processes taking place in the horizon's neighborhood. In particular, an astronomical observer far away from the black hole measures electromagnetic (or gravitational) radiation coming from the near region of the black hole. This emitted radiation follows outgoing null geodesics and the astronomical  observer receives the radiation at the asymptotic end of the  outgoing null hypersurfaces generated by those geodesics. 

Mathematically there are two ways to asymptotically analyze radiation fields; in the first approach,   matter  fields and the  physical metric $g_{ab}$ are expanded in the physical spacetime with respect to inverse powers of a suitable  radial coordinate while the  second approach employs the so-called Penrose compactification of spacetime \cite{Penrose1963}. This compactification consists in attaching a null boundary to the physical spacetime. Thereby an extended conformal spacetime manifold is built using a conformal metric $\hat g_{ab}=\Omega^2 g_{ab}$ in which $\Omega$ is a suitable conformal factor vanishing at the null boundary. The attached null boundary is called null infinity, ${\mathscr{I}}$, and a local Taylor series expansion of geometrical and physical quantities off $\mathscr{I}$  in the  conformal spacetime allows one to  mathematically  analyse the radiation fields. There are in fact two such boundaries $\mathscr{I}^+$ and $\mathscr{I}^-$, also known as future null infinity and past null infinity.  Indeed, an idealized astronomical observer 
would be placed at $\mathscr{I}^+$, in the far future of the black holes.

The first convincing understanding of nonlinear radiation fields in general relativity has been done by Bondi and collaborators \cite{Bondi, Sachs}. They introduced a chart consisting of an (outgoing) null coordinate $u$ (corresponding to the retarded time in Minkowski spacetime), an areal distance $r$ and two spherical angles $x^A = (\theta,\phi)$. Furthermore they required the metric to approach a Minkowski metric in outgoing polar null coordinates for the (physical) spacetime metric $g_{ab}$ that is expanded in inverse powers of $r$. In the asymptotic region $u=const$ are null hypersurfaces, whose generating rays are parameterized with $r$. 
This Bondi null coordinate $u$ however is not well suited to study fields at the horizon of a black hole. As an example, we consider  again charts in Schwarzschild spacetime. First, with the well known tortoise coordinate $r^*(r)$, the outgoing Eddington-Finkelstein coordinate $u = t-r^*$  takes the form  $u=t-r$ for large values of $r$, where $t$ is the inertial  time of the asymptotic observer, but $u$ is singular at the horizon $r=2m$.  Second, Kruskal-Szekeres coordinates are well defined at the horizon and they allow us to  understand the conformal structure of the Schwarzschild spacetime; 
however, the coordinates  have the caveat that the areal distance coordinate $r$ is expressed as an implicit function in terms of the Kruskal's null coordinates.  and the standard flat space null coordinates. Because of this the analysis of fields near the horizon and at large distances in his chart  is difficult. But there is another (less known) global representation of the Schwarzschild spacetime due to Israel \cite{Israel1966,Israel1967} ( and rediscovered by \cite{Pajerski1971, Kloesch1996}, see also Blau's online lecture notes for a complete discussion\cite{Blau}) where the metric of a Schwarzschild black hole takes a simple and explicit form with rational functions
\begin{equation}\label{BH_israel}
\begin{split}
g_{ab}dx^adx^b =& -\frac{2y^2}{8m^2 -wy}dw^2 +2dwdy \\
&- 
\Big(2 m -\frac{wy}{4m}\Big)^2 (d\theta^2+\sin^2\theta d\phi^2)\\
\end{split}
\end{equation}
Israel obtained this metric by analyzing the null geodesics in the standard Schwarzschild metric representation adopting the $w$ coordinate to the null structure. The past and future horizons in the above metric are given by $y=0$ and $w=0$, respectively. Note that the radial coordinate $x^1= y$ is an affine parameter of the null vectors generating the null hypersurfaces $w=const$, which is indicated by $g_{wy}=1$. For an asymptotic analysis using Penrose' compactifications scheme,  introduce an inverse affine parameter $\Upsilon=(4m)/y$, a rescaling $ w \rightarrow 4mw$ and a conformal factor $\Omega = \Upsilon/(4m)$, we then discover that the metric is given by the  conformal metric
\begin{equation}\label{metricOffScri}
\begin{split}
&\hat g_{ab}dx^ad x^b = \Omega^2 g_{ab}d\hat x^ad\hat x^b
\\&=
-2dwd{\Upsilon} - w^2(d\theta^2+\sin^2\theta d\phi^2) + O({\ell})
\end{split}
\end{equation}
which is well defined for ${\Upsilon}=0$, i.e. at null infinity. We can see that the  coordinate pair  $(w,y)$ consists of  bona-fide coordinates so that it allows us to construct a coordinate chart at the horizon $y=0$ as well as in the asymptotic region for $y\rightarrow \infty$. It is of interest to see whether such pair $(w,y)$ exist in a general sense, so that it can be used to chart black hole spacetimes from a horizon to null infinity. The main purpose of this work is to give  an affirmative answer for various spherically symmetric spacetimes. Thereby we present two possible scenarios for achieving this aim. In the first one we follow previous works of \cite{Kozameh:2012pw,Kozameh11}, where a regular null coordinate is constructed based on geometrical restrictions (see Sec.~\ref{sec:reg_coord_setup}). In the second approach, we follow the affine-null metric formulation of Einstein equations \cite{Win2013, TM2019, Crespo2019}, which is a formulation of Einstein equations with respect to an affine-null metric. This formulation shares similarities with  the Bondi-Sachs metric approach of General  Relativity in which the relevant field equations are cast into a hierarchical system \cite{Scholarpedia}. We demonstrate at the example of the Reissner-Nordstr\"om solution that the two approaches lead to equivalent results.

The regular null coordinate framework of \cite{Kozameh:2012pw,Kozameh11} is summarized in Sec.~\ref{sec:reg_coord_setup}. The following sections employ this framework for non-extremal (Sec.~\ref{non_ex}) and extremal spherically symmetric black holes (Sec. \ref{ex}). We also show how it can be used to find a regular null coordinate version of the outgoing (Sec.~\ref{vai_out}) and ingoing (Sec.~\ref{vai_in}) Vaidya solution. The spherically symmetric affine null metric formulation is discussed in Sec.\ref{affnull}, where the Reissner-Nordstr\"om solution is derived using for the first time a characteristic initial value formulation to obtains the charged version of  the metric \eqref{BH_israel}. 

 To be in lines with  \cite{Kozameh:2012pw,Kozameh11}, we make use of   the notation of Geroch, Held and Penrose (GHP) \cite{Geroch73}, and we work exclusively in with the negative  signature convention $-2$ for the physical metric.

\begin{figure}[h]
\includegraphics[clip,width=0.5\textwidth]{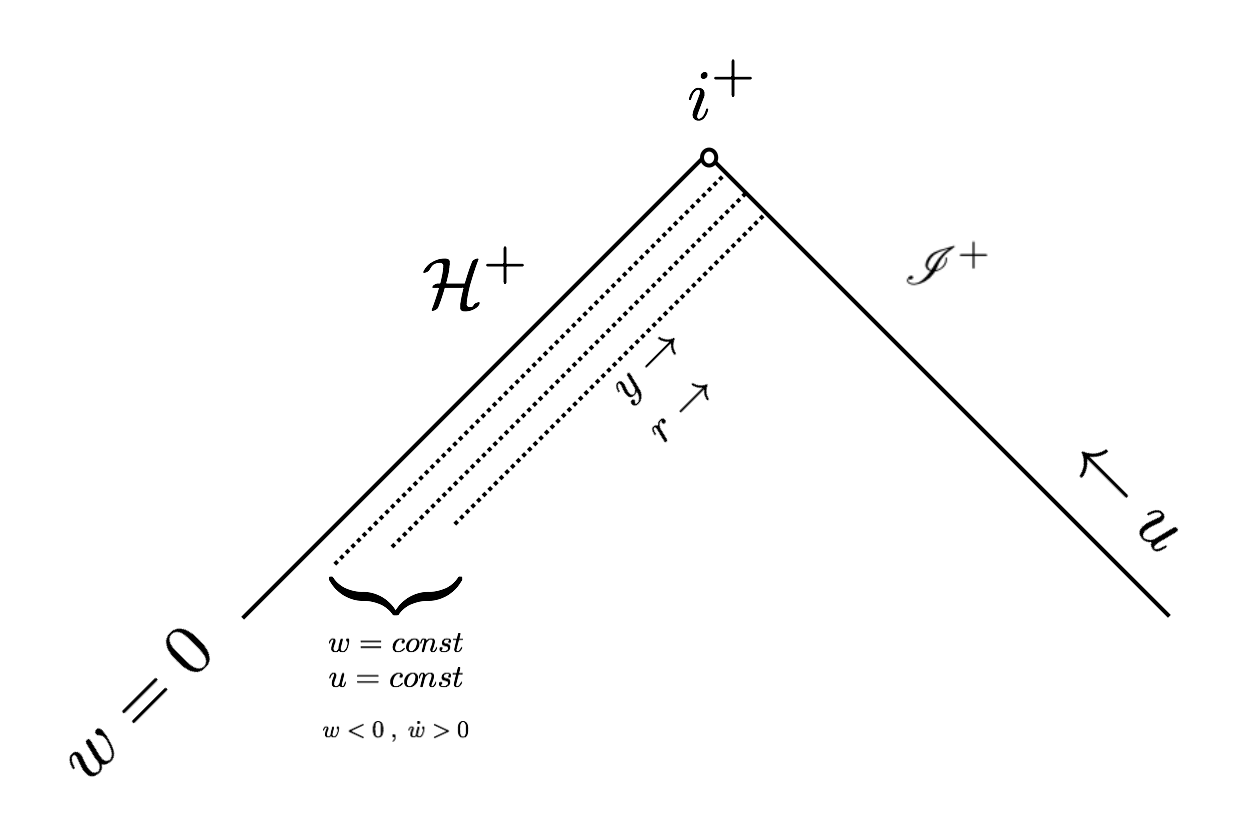}
\caption{
In the past of an open
set of future null infinity defined by those points
for which their Bondi
retarded time u is in the range
$u\in (u_0,\infty)$ we require the existence of a regular null
function $w$ such that: $w = 0$ at the horizon $\mathcal{H}^+$, and $w < 0$
in the region of interest.  
 { In a vicinity of $i^+$, the null surfaces of constant retarded time $u$
are smooth all the way up to the event horizon for SBHs.}
}
\label{figui}
\end{figure}

\section{The framework}\label{sec:reg_coord_setup}

In the mentioned framework regular
null coordinates are introduced based on the assumption that a suitable family of null surfaces are caustic free in a neighbourhood of timelike infinity $i^+$ containing a portion of the
black hole horizon $\mathcal{H}^+$ and future null infinity $\mathscr{I}^+$.

In this paper, we particularize to
spherically and asymptotically flat spacetimes at future null
infinity $(M, g_{ab})$ containing a black hole; for more details about the formalism we refers to \cite{Kozameh:2012pw,Kozameh11}. 

We choose a Bondi coordinate $u$ in such a way that it coincides with the center of mass Bondi cuts in the regime $u\to\infty$ limit.  In the past of an open
set of future null infinity ($\mathscr{I}^+$) defined by those points for which their Bondi retarded time $u$ is in the range $u\in(u_0,\infty)$  we require there exists a smooth null
function $w=w(u)$ such that $w = 0$
at the horizon $\mathcal{H}^+$, $\lim_{u\to\infty}w = 0$, $\dot{w}\equiv\frac{dw}{du} > 0$, and $w < 0$ for all $u$ in the (exterior) region between $\mathcal{H}^+$ and $\mathscr{I}^+$. 
The kind of metrics satisfying these conditions are referred as solitary black holes (SBHs) in \cite{Kozameh:2012pw,Kozameh11}.
The associated conformal diagram is depicted in Figure 1. 

The null geodesic
congruence defined by $\tilde\ell = du$ allows for the introduction of an affine parameter $r$ used as a radial coordinate
which is fixed by the requirement that it coincides asymptotically with the luminosity distance\cite{Kozameh:2012pw}. The surfaces $(r, u)$=constant are spheres which inherit natural
spherical coordinates defined in the Bondi cuts at $\mathscr{I}^+$
which label null rays of the congruence. All this provides a coordinate system $(u, r, x^A)$ with $x^A$ coordinates of the two-dimensional sphere $S^2$. 
In a similar way, we define the one form $\ell_a\equiv (d w)_a$, then,
the geodesic vector field $\ell^a$  is also tangent to the
null congruence defined by $\tilde\ell^a$. It is therefore
natural to introduce the affine function $y$ through
$\ell^a=\left(\frac{\partial}{\partial y}\right)^a$. Therefore, the functions $(w,y)$ can be used as coordinate functions in the region
where  the null congruence $\ell^a$ does no show caustics.
The affine
parameter $y$, for each null geodesic, can be choosed 
so that the 2-spheres $u$=const., $r$=const. coincide with the
 2-spheres  $w$=const., $y$=const, implying the following relationship between $r$ and $y$:
\be
r=\dot w y + r_{0}(w)
\label{eqq}.
\ee
Hence, one has a new coordinate system $(w, y, x^A)$ where $x^A$ are again coordinates of $S^2$.
We assume that $r$ is a smooth function of $(w,y)$
all the way up to the horizon.

Let us observe that
from the null vector fields $\ell^a$ and $ \tilde \ell^a$ one can construct null tetrads
$(\ell^a,m^a, \bar m^a, n^a)$, and $(\tilde \ell^a, \tilde m^a, \bar{\tilde m}^a,\tilde n^a)$ adapted to the
geometry of the coordinate system introduced above.\footnote{
With the usual normalization  $1 = \ell^a n^{}_a = - m^a \bar m^{}_a$ and $1 = \tilde\ell^a \tilde n_a = - \tilde m^a \bar{\tilde m}_a$
with all other respective scalar products being zero.} The freedom in this choice is reduced by choosing  the vectors $m^a= \tilde m^a$,
and tangent to the topological 2-spheres $(w,y)=\text{constant}=(u,r)$.

From $w=w(u)$ it follows that $dw=\dot w \,du$ which implies the following relation between the two tetrads
\be\ell^a=\dot w\  \tilde \ell^a, \ \ \  n^a=\frac{ 1}{\dot w}\tilde n^a, \ \ \ m^a= \tilde m^a. \label{mainy} \ee
If we denote the five (complex) Weyl tensor spinor components\cite{Geroch73}  $\Psi_{N}^{}$ and $\tilde \Psi_{N}^{}$ for $N\in \{0,1,2,3,4\}$
in each of the respective tetrads, then we get the following relations for the Weyl curvature scalars $\Psi^{}_N = \dot w^{(2-N)}  \tilde \Psi^{}_N$.
Similar relations are obtained for the Ricci curvature scalars, in particular 
$\Phi^{}_{00}= \dot w^{2}  \tilde \Phi^{}_{00}$, $\Phi^{}_{11}= \tilde \Phi^{}_{11}$ and $\Phi^{}_{22}= \dot w^{-2}  \tilde \Phi^{}_{22}$.
 
The above assumptions of regularity of the new coordinates at the horizon imply that the limit
 $r_H \equiv \lim_{w\rightarrow 0} r(w,y)$ exists and is constant\cite{Kozameh:2012pw,Kozameh11}.

 Since by assumption $\dot w(w)$ admits a Taylor expansion around $w=0$ we can write:
\be \label{eq:wexpans}
\dot w =a(w)=a_1 w+ \sO(w^2).
\ee
 Assuming that $a_1\not=0$ the above equation can be integrated giving the
important relation
\be\label{main}
{
w(u)=-\exp{(a_1(u-u_0))}+\sO(\exp{(2a_1u)}) }  ,
\ee
where $\exp(-a_1 u_0)$ is the rescaling freedom mentioned previously associated with the choice of origin for
the Bondi retarded time $u$.

As shown in \cite{Kozameh11} $a_1$ has
a clear geometrical meaning as follows from the properties of the vector field $\chi = \partial_u$: 

\begin{enumerate}
 \item
It is a smooth vector field that is a null geodesic generator at $\sI^{+}$. As $u$ is a Bondi coordinate
it generates inertial time translations at future null infinity.

 \item
It is a null geodesic generator of the horizon $\mathcal{H}^+$.

 \item
At the horizon $\mathcal{H}^+$, $\chi$ satisfies the equation,
\begin{equation}\nonumber
\chi^{a}\nabla_{a}\chi^{b} \equiv k_H \chi^{b} ;
\end{equation}
where $k_H$ is a  generalized surface gravity.

\item The coefficient $a_1$ is the negative of the surface gravity $k_H$, i.e. 
\be{ a_1=-k_H = \text{const.} } \label{kappa}\ee

\end{enumerate}
For a proof of these properties, we refer to \cite{Kozameh11}.

Therefore, the family of spacetimes considered here admits  a notion of surface gravity which
coincides with the usual one in cases when the
spacetime is stationary.
Note that
if we had taken $a_1=0$ above, one would have obtained $k_{H}=0$. This situation corresponds to the especial cases
involving (in particular) the stationary
extremal black holes. We will discuss this case in Sec.\eqref{ex}.

With this definition of surface gravity, the relation between
the null coordinate $w$ and the Bondi retarded time $u$ in the case of $k_H\neq 0$ reads%
\begin{equation}\label{eq:expdecay}
w=-\exp{(-k_{H}(u-u_{0}))}+{\mathfs {O}}(\exp{(-2k_{H}u)})
.
\end{equation}

\section{The formalism applied to spherical symmetric black holes}
In this section, we specify the regular coordinates $(w,y)$ of the  previously explained formalism for static spherically symmetric  spacetimes. 
We discuss the non-extremal and extremal solutions. 


Consider a static and spherically symmetric metric in Bondi coordiantes with timelike Killing vector $\partial_u$,
\begin{equation}\label{eq:metricbondifr}
ds^2=f(r)du^2+2dudr-h(r)^2d\Omega^2, 
\end{equation}
where $d\Omega^2=d\theta^2+\sin^2\theta d\phi^2,$ is the metric of an unit sphere written in standard spherical coordinates, $(\theta,\phi)$, and $h(r)$ a smooth function which is non-vanishing at the horizon.
The event horizon of this metric is located on the null surface $r=\text{constant}$ 
placed in the bigger root of $f(r)=0$. We will call this root $r_H$ and assume that $f(r)$ and $h(r)$  are regular functions at $r=r_H$.
A null tetrad adapted to this coordinate system is given by:
\begin{eqnarray}
\tilde{\ell}^a&=&\left(\frac{\partial}{\partial r}\right)^a,\\
\tilde{n}^a&=&\left(\frac{\partial}{\partial u}\right)^a-\frac{f(r)}{2}\left(\frac{\partial}{\partial r}\right)^a,\\
\tilde{m}^a&=&\frac{1}{\sqrt{2}h(r)}\left[\left(\frac{\partial}{\partial \theta}\right)^a
+\frac{i}{\sin(\theta)}\left(\frac{\partial}{\partial \phi}\right)^a\right],\\
\tilde{\bar{m}}^a&=&\frac{1}{\sqrt{2}h(r)}\left[\left(\frac{\partial}{\partial \theta}\right)^a
-\frac{i}{\sin(\theta)}\left(\frac{\partial}{\partial \phi}\right)^a\right].
\end{eqnarray}
Using the GHP notation\cite{Geroch73} the resulting expressions for the non vanishing spin coefficients are: 
\begin{equation}
\begin{split}
\tilde{\rho}&= -\frac{1}{h}\frac{dh}{dr},\;\;,\;\;
\tilde{\rho}'= \frac{f}{2h}\frac{dh}{dr},\\
\tilde{\epsilon}'&=-\frac{1}{4}\frac{df}{dr},\;\;,\;\;
\tilde{\beta} =\tilde{\beta}'=\frac{\sqrt{2}}{4h}\cot(\theta).
\end{split}
\end{equation}
In particular, as it is well known, the expansion $\tilde{\rho}$ associated 
to the null congruence $\tilde{l}^a$ is not vanishing if  
evaluated at the horizon.
This unphysical result is a consequence that this 
coordinate system is not regular there.
\subsection{The non-extremal case ($k_H\neq 0$)}  
\label{non_ex}

Now, we wish to make a coordinate transformation to regular coordinates on the horizon $(w,y)$.
To do that we assume that the black hole of interest is non-extremal, i.e. the case where
\be 
\dot w=a_1 w+O(w^2).
\ee Moreover, we assume that the relation between $\dot w$ and $w$ is exactly linear, giving 
\begin{eqnarray}
w&=&e^{a_1u},\label{eq:relwu}\\
r&=&a_1w y+r_H\label{eq:relry}.
\end{eqnarray}
The extremal case will be dealt with below. 
The previous equations imply 
\begin{eqnarray}
du&=&\frac{dw}{a_1w},\\
dr&=&a_1ydw+aw\,dy.
\end{eqnarray}
By replacing these relations into eq.(\ref{eq:metricbondifr}), we get
\begin{equation}\label{eq:metricsphegen}
\begin{split}
ds^2&=\left(\frac{f(a_1w y+r_H)}{a_1^2w^2}+\frac{2y}{w}\right)dw^2+2dw\,dy\\
&-[h(a_1w\,y+r_H)]^2d\Omega^2. 
\end{split}
\end{equation}
This is an exact expression for the class of metrics (\ref{eq:metricbondifr}) 
in terms of regular coordinates $(w,y)$. As follows from eq.\eqref{eq:relry}, in these coordinates the horizon $\mathcal{H^+}$ is placed at $w=0$.

We would like to study these 
metrics in the limit $w\rightarrow 0$ at a neighborhood of the horizon. As of the regularity requirements,
we can make an expansion of  $g_{ww}$ in terms of $w$. In a
 neighborhood of the horizon of size $a_1yw\ll r_H$ we have,
\begin{equation}\label{eq:taylf}
\begin{split}
f(r_H+a_1yw)=&f(r_H)+f'(r_H)a_1yw\\&+\frac{1}{2}f''(r_H)a^2_1y^2w^2+
\textit{O}(w^{3})\\
=&f'(r_H)a_1yw+\frac{1}{2}f''(r_H)a^2_1y^2w^2+\textit{O}(w^{3}),
\end{split}
\end{equation}
where we used  
$f(r_H)=0$. Note that the regularity of the metric \eqref{eq:metricsphegen} at $w=0$ requires
\begin{equation}\label{eq:sssg}
a_1=-k_H=-\frac{f'(r_H)}{2}.
\end{equation}
which is true for static and spherically symmetric metrics. This is an independent 
way of proving \eqref{kappa} which is valid in the spherically symmetric situation we are considering.
 Hence, the expansion in eq.\eqref{eq:taylf} becomes
\begin{equation}\label{eq:taylf-bis}
\begin{split}
f(r_H+a_1yw)
=&-2a^2_1y w+\frac{1}{2}f''(r_H)a^2_1y^2w^2+\textit{O}(w^{3}).
\end{split}
\end{equation}

A similar expansion follows for $h(r)$, but starting with a non-zero constant term $h(r_H)$ (because the area of the horizon is not zero).

Therefore by replacing this expression into eq.(\ref{eq:metricsphegen}) we get
\begin{equation}
\begin{split}
g_{ww}=&\frac{1}{2}f''(r_H)y^2+\frac{1}{6}f'''(r_H)a_1w\,y^3+\cdots\\
&+\frac{1}{n!}f^{(n)}(r_H)(a_1w)^{n-2}y^n+\textit{O}(w^{n-1}).\label{eq:gww}
\end{split}
\end{equation}
In particular the 4-dimensional spacetime metric at a vicinity of the horizon reads,
\ba 
\left.ds^2\right|_{r=r_H}&=&\frac{1}{2}f''(r_H)y^2dw^2+2dw\,dy-h^2(r_H)d\Omega^2\n \\  &+&\textit{O}(w^{3}).\label{eq:ds2horizon}
\ea

The spin coefficients associated to the new null coordinate system are:
\begin{eqnarray}
\rho&=&\dot{w}\tilde{\rho}=-\dot{w}\left.\left[\frac{1}{h}\frac{dh}{dr}\right]
\right|_{r=\dot{w}y+r_H},\\
\rho'&=&\frac{1}{\dot{w}}\tilde{\rho}'=\frac{1}{2\dot{w}}\left.\left[\frac{f}{h}\frac{dh}{dr}
\right]\right |_{r=\dot{w}y+r_H},\\
\epsilon'&=&\left.\left[\frac{\tilde{\epsilon}'}{\dot{w}}-\frac{1}{2}\tilde{n}^a\nabla_a(\dot{w}^{-1})\right]
\right|_{r=\dot{w}y+r_H}\nonumber\\
&=&
\left.\left[-\frac{1}{4\dot{w}}\frac{df}{dr}+\frac{k_H}{2\dot{w}}\right]\right|_{r=\dot{w}y+r_H},\\
\beta&=&\tilde{\beta}=\left.\left[\frac{\sqrt{2}}{4h}\cot(\theta)\right]
\right|_{r=\dot{w}y+r_H}.
\end{eqnarray}
Note in particular, that now $\rho=0$ at the horizon (as it should be), and even when $\rho'$ and $\epsilon'$ 
have a factor $\dot{w}^{-1}$
in their expressions, they are regular at the horizon as it can be seen by studying
their limit using  eq.(\ref{eq:taylf}) and eq.(\ref{eq:sssg}).
For completeness we give expressions for the non-vanishing scalar curvatures,
\begin{eqnarray}
\Phi_{00}&:=&\frac{\dot\omega^2}{h}\frac{d^2h}{dr^2},\\
\Phi_{11}&=&-\frac{1}{8h^2}\left[\frac{d^2f}{dr^2}h^2+2-2f\Big(\frac{dh}{dr}\Big)^2\right],\\
\Phi_{22}&=&\frac{f^2}{4\dot\omega^2 h}\frac{d^2h}{dr^2},\\
\Lambda&=&-\frac{1}{24h^2}\left[\frac{d^2f}{dr^2}h^2+4h\frac{dh}{dr}\frac{df}{dr}+4fh\frac{d^2h}{dr^2}\right.\nonumber\\
&&\qquad\left.-2+2f\Big(\frac{dh}{dr}\Big)^2\right],\\
\Psi_2&=&\frac{1}{12h^2}\left[\frac{d^2f}{dr^2}h^2-2fh\frac{d^2h}{dr^2}-2h\frac{dh}{dr}\frac{df}{dr}
\right.\nonumber\\
&&\qquad\left.-2+2f\Big(\frac{dh}{dr}\Big)^2\right].
\end{eqnarray}

Note that taking into account the eq.\eqref{eq:taylf-bis} all these curvature scalars are regular at $w=0$, including $\Phi_{22}$.

As an example let us consider a Reissner-Nordstr\"om black hole, \begin{eqnarray}
f(r)&=&1-\frac{2m}{r}+\frac{Q^2}{r^2},\\ 
h(r)&=&r,\\ r_H&=&m+(m^2-Q^2)^{\frac{1}{2}},\\
k_H&=&\frac{r^2_H-Q^2}{2r^3_H}
\end{eqnarray}
we get  from \eqref{eq:metricsphegen}
\begin{equation}
ds^2=\frac{(1-2k_H(r+r_H))y^2}{r^2}dw^2+2dw\,dy-r^2d\Omega^2.\label{eq:rgene}
\end{equation}
with $r=-k_Hwy+r_H$.
In these coordindates, $y=0$ corresponds to the non-expanding null hypersurface ${\cal H}^-$ being the the past null horizon. The future horizon ${\cal H}^+$ is at $w=0$.

In the case $Q=0$, we obtain the Schwarzschild solution
\begin{equation}
ds^2=-\frac{2k_Hy^2}{-k_Hwy+2m}dw^2+2dw\,dy-(r_H-k_Hw\,y)^2d\Omega^2.\label{eq:rgene}
\end{equation}
with corrresponding $k_H = (2r_H)^{-1}$.
 This is exactly the solution found by Israel (see eq.\eqref{BH_israel}) \cite{Israel1966,Israel1967} (and rediscovered by  Pajerski and Newman \cite{Pajerski1971}  as well as Kl\"obsch and  Strobl \cite{Kloesch1996}, who also obtained the Reissner-Norstr\"{o}m metric in these coordinates using a connection between this metric and highly symmetric solutions of particular two-dimensional generalized dilaton gravity models.)

 As noted by Blau \cite{Blau}, the Schwarzschild metric as expressed in eq.\eqref{eq:rgene} admits the isometry $\tilde{w}=\lambda w$, $\tilde{y}=\lambda^{-1}y$ which corresponds to the timelike Killing vector $\chi=-k_H\left(w\partial_w-y\partial_y\right)$. This property is also shared by the more general metric given by eq.\eqref{eq:metricsphegen}. 
 
 As a final remark, let us note that our construction of regular coordinates $\{w,y\}$ for static and spherically symmetric asymptotically flat spacetimes  is not restricted to metrics which are solutions of the Einstein's equations, as long as the assumptions given by the regularity requirements  are satisfied. 
\subsection{The extremal case $k_H=0$}\label{ex}

In many situations, for certain choices of parameters that describe a black hole, an extremal solution can be obtained where the surface gravity is zero. Let us consider for example the extremal case with $f=(1-\frac{r_H}{r})^2$. An extremal Reissner-Nordstr\"om solution with $Q=m$ falls into this family. In such a case, as the surface gravity is zero, (and therefore $a_1=0$) a naive ansatz for eq.\eqref{eq:wexpans} would be to consider that $\dot{w}=a_2w^2$, however it can be checked that the resulting expression for the metric in $\{w,y\}$ coordinates is not regular at $w=0$. In the next subsections we will present two alternative approaches to solve this problem.

\subsubsection{Approach I: The direct construction of an analytic null function $w$}
The extreme Reissner-Nordström metric can be expressed\cite{Chandra} by:
\begin{equation}\label{eq:r-d}
\begin{split}
ds^2 =& 
f(r) dt^2
-\frac{dr^2}{f(r)} - r^2 d\Omega^2
;
\end{split}
\end{equation}
where 
\begin{equation}
f(r)=\left(1-\frac{m}{r}\right)^2.
\end{equation}

It is customary to define the tortoise coordinate
\begin{equation}
r^* = \int_{r_1}^{r} \frac{dr}{f}
;
\end{equation}
which for the extremal Reissner-Nordström metric gives
\begin{equation}
r^* = 
r + 2 m \ln\big(\frac{r - m}{m}\big)
- \frac{m^2}{r-m} + c_1
.
\end{equation}
Then, it is usual to define the (outgoing/ingoing) null coordinates,
\begin{equation}\label{eq:u}
u = t - r^* ,
\end{equation}
and
\begin{equation}\label{eq:v}
v = t + r^* .
\end{equation}

In general one can consider other families of null coordinates
where the metric can be expressed as:
\begin{equation}\label{eq:r-d-1}
\begin{split}
ds^2 =& 
-4 f(r) A(u) B(v) du dv - r^2 d\Omega^2
,
\end{split}
\end{equation}
with
\begin{equation}\label{key12}
dr = 
-f(r)
\bigg( A(u) du + B(v) dv \bigg)
;
\end{equation}
which for the previous case of \eqref{eq:u} and \eqref{eq:v} one has to take
\begin{equation}\label{key13}
A(u) = -B(v) = \frac{1}{2}
;
\end{equation}
and note that in this case one has
\begin{equation}\label{key1}
dr^* = \frac{1}{f} dr 
= -\frac{1}{2} (du - dv)
.
\end{equation}

We now look for a null coordinate $w$ which is regular
near the horizon, with
\begin{equation}\label{eq:r-d-2}
\begin{split}
ds^2 =& 
-f(r)A(w) B(v) dw dv - r^2 d\Omega^2
.
\end{split}
\end{equation}

Recall that the null radial geodesic equation 
is~\cite{Chandra}(p.216,eq.70):
\begin{equation}
\frac{dr}{d\lambda}
=
\pm E ;	
\end{equation}
so that along these null geodesics the radial coordinates
is proportional to the affine parameters.
In particular for the incoming null radial geodesics one
has
\begin{equation}\label{eq:rel}
\frac{dr}{d\lambda}
=
- E(v) ;	
\end{equation}
where we have the freedom to choose for different $v$'s
different constants $E>0$.

Let us consider then the incoming null geodesics, that is
with $dv=0$.
Then, in the integral form one must have
\begin{equation}\label{key2}
r^* = 
-\int \alpha'(w) dw ;
\end{equation}
where $\alpha'(w) \equiv \frac{d\alpha}{dw}$; so that
\begin{equation}\label{key3}
r^* =
r + 2 m \ln\big(\frac{r - m}{m}\big)
- \frac{m^2}{r-m} + c_1
= - \alpha(w)
.
\end{equation}

{Since $r$ behaves as the affine parameter, we can think of
$\lambda$ as given by}
$E(v) \lambda = -{(r - m)}$,
and as we want $w$ to be regular at the horizon, $r=m$,
we take, along a null geodesic in $v=v_0$, $E(v_0)=E_0>0$ and
$w=\lambda$, so that we set
\begin{equation}\label{eq:alfa}
\frac{\alpha}{m}
=
- 2 \ln(-\frac{E_0}{m}w) - \frac{m}{E_0w} + \frac{E_0}{m}w
;
\end{equation}
since in this way we capture the two terms
with divergent behaviors, and
where we have divided by $m$ the original expression
to deal with quantities without units.

 {Then, recalling the relation between $r^*$
and $u$ at constant $v$, we have}
\begin{equation}\label{keyo1}
\frac{\alpha}{m}
=
- 2 \ln(-\frac{E_0}{m}w) - \frac{m}{E_0w} + \frac{E_0}{m}w
= \frac{u - u_0}{2 m}
;
\end{equation}
{where without loss of generality we can take $u_0=0$,}
so that after differentiation with respect to $u$ we obtain
\begin{equation}\label{keywd}
\dot{w}=\frac{1}{2}\frac{E_0w^2}{(E_0w-m)^2}.
\end{equation}
with a Taylor expansion around $w=0$,
\begin{equation}\label{eq:wdoto}
\dot{w}
=
\frac{E_0}{2 m^2} (w^2 + 2\frac{E_0}{m} w^3) +\mathcal{O}(w^4)
.
\end{equation}
Note that $w=0$ correspond to $u\to\infty$, $\dot{w}>0$ in the exterior region of the black hole and it has an analytic expansion in powers of $w$. Equation \eqref{keyo1} (or equivalently \eqref{keywd})  define our desired $w$ coordinate.\\

{
\paragraph{Relation with coordinate $y$. }
With respect to the coordinate $y$, for an incoming null geodesic,
contained in the hypersurface $v=v_0$, one will have a functional
dependence of the form $y_v(w)=y_v(\lambda)$.

The general relation between $r$ and $y$ is of the form:
\begin{equation}\label{key4}
r = \dot{w} \big(y - y_0(w)\big) + \tilde{r}_0(w) 
= \dot{w} y + {r}_0(w)
;
\end{equation}
where we know that in general $r_H \equiv r_0(w=0)$ is the radius
of the horizon.

Then, along the incoming null geodesic,
contained in the hypersurface $v=v_0$, one will have
\begin{equation}\label{eq:rdey}
r = \dot{w} (y_v(w) - y_0(w)) + \tilde{r}_0(w) = m - E_0w\,  ;
\end{equation}
due to the previous relation between affine parameter
and radial coordinate.

Let us note that at this stage we have the freedom to chose
$y_0(w)$ and the function $\tilde{r}_1(w)$, in $\tilde{r}_0(w) = r_H + \tilde{r}_1(w)$,
with $\lim\limits_{w\rightarrow 0}\tilde{r}_1(w) = 0$, and $r_H = m$.

\paragraph*{{\bf Choice a)}}
Let us consider the choice of $y_0(w)$ so that:
\begin{equation}\label{key5}
\dot{w} y_0(w) = \tilde{r}_1(w) ;
\end{equation}
then we would have
\begin{equation}\label{key22}
r = \dot{w} y  + r_H ;
\end{equation}
so that when we take $y=0$ one would have
$r = r_H$, that is the past horizon $\mathcal{H}^-$.
Note that in this case $y=0$ implies an incoming null
geodesic contained in $\mathcal{H}^-$. 

Let us note that although this seems to be a natural
choice, it might involve a singular definition for
the coordinate $y$; which is related to the fact that $\mathcal{H}^-$ can not be taken as the initial incoming null hypersurface, 
used in the previous mechanism to define the coordinate $w$, 
since it is outside the manifold covered by $v$.

\paragraph*{{\bf Choice b)}}
From the previous discussion one is tempted to consider
\begin{equation}\label{key6}
\tilde{r}_1(w) = - E_0w.
\end{equation}
Then, for incoming null geodesic,
contained in the hypersurface $v=v_0$, one will have 
from \eqref{eq:rdey} that
\begin{equation}\label{eq:y_v}
\dot{w} (y_v(w) - y_0(w)) = 0
 ;
\end{equation}
so that in particular, by taking $y_0(w) = 0$
one would have that at the original
incoming null geodesic 
\begin{equation}\label{key7}
y_v(w) = 0 ;
\end{equation}
that is we choose in this way that the value 0
of coordinate $y$ is at the original incoming
null geodesics.
Then we would have
\begin{equation}\label{eq:r0maest}
    r=\dot{w}y+r_H-E_0w.
\end{equation}
Note that for {\bf Choice a)} one has ${r}_0 = r_H = m$;
while  {\bf Choice b)} one has 
${r}_0 = r_H - E_0w = m - E_0w$.
}
\\
{
\paragraph{Relation with the $W:=g_{ww}$ component.}\label{sec:d}

Using $dw = \dot{w} du$, one has
\begin{equation}\label{key8}
f  du^2 + 2 du dr
=
\frac{f}{\dot{w}^2} dw^2 + \frac{2}{\dot{w}} dw dr-r^2d\Omega^2
. 
\end{equation}
And using the relation of $r(w,y)$ one has
\begin{equation}\label{key9}
\begin{split}
ds^2
=&
\bigg(
\frac{f}{\dot{w}^2}
+ \frac{2}{\dot{w}} \big(\frac{d \dot{w}}{dw} y + \frac{d {r}_0}{dw}\big)
\bigg)dw^2
+ 2 dw dy-r^2d\Omega^2
;
\end{split}
\end{equation}
which shows the $g_{ww}$ dependence on the choice
of gauges $y_0$ and $r_0$ or equivalently $\tilde{r}_0$.
Note that
\begin{equation}\label{key10}
\begin{split}
W=&
\frac{(\dot{w} y + {r}_0(w) - m)^2}{\dot{w}^2(\dot{w} y + {r}_0(w))^2}
+ 
\frac{2}{\dot{w}}
\big(
\frac{E_0mw}{(m-E_0w)^3}
y + \frac{d{r}_0}{dw}
\big)
;
\end{split}
\end{equation}
so that for {\bf Choice a)} one has
\begin{equation}\label{keyWnoaf}
\begin{split}
W =&
\frac{ y^2}{(\dot{w} y + m )^2}
+ 
\frac{2}{\dot{w}}
\frac{E_0mw}{(m-E_0w)^3} y
.
\end{split}
\end{equation}
However, it has not a well behaviour at $\mathcal{H}^+$, $W=4y/w+y^2/m^2+4E_0y/m+\mathcal{O}(w)$. Therefore, it does not serve our purposes.

Notwithstanding, for {\bf Choice b)} one has
\begin{widetext}
\begin{equation}\label{key11}
\begin{split}
W=&
\frac{(\dot{w} y - E_0w )^2 }{\dot{w}^2(\dot{w} y + m - E_0w)^2}
+ 
\frac{2}{\dot{w}}
\big(
\frac{E_0mw}{(m-E_0w)^3} y 
- E_0
\big)
\\
=&-\frac{4(2w^4E_0^4-5w^3E_0^3m+3w^2E_0^2m^2+wE_0m^3-w^3yE_0^2-m^4)my^2}{(-E_0w+m)(-2w^3E_0^3+6w^2E_0^2m-6wE_0m^2+2m^3+w^2E_0y)^2};
\end{split}
\end{equation}
\end{widetext}
which has a well behaviour at a neighborhood of  $\mathcal{H}^+$, $W\approx y^2/m^2+6y^2E_0w/m^3+\mathcal{O}(w^2)$. Note also that $W|_{y=0}=W_{,y}|_{y=0}=0$, in agreement with the condition that the hypersurface $y=0$ be null with $w$ as an affine parameter of its null vector generator $n=\partial_w$ (cf. Sec.\eqref{affnull}). 

Hence, the coordinate system $(w,y)$ related to the Bondi coordinates $(u,r)$ by
\begin{eqnarray}
u&=&2m(- 2\ln(-\frac{E_0}{m}w) - \frac{m}{E_0w} + \frac{E_0}{m}w),\\
r&=&\frac{1}{2}\frac{E_0w^2}{(E_0w-m)^2}y+m-E_0w;\label{q:rfinalwog}
\end{eqnarray}
is the kind of coordinates that accomplishes the requirements of Sec.\eqref{sec:reg_coord_setup}.
}

\subsubsection{Approach II: Requiring regularity of $W$ at the horizon}
{
 We can also consider a more general transformation and see the necessary conditions in order the metric be regular at $w=0$. First, let as note that at the considered case $f(r_H)=f'(r_H)=0$.
Le us assume that 
\begin{equation}\label{eqr0}
    r=\dot{w}y+r_0(w)=a(w)y+r_0(w),
\end{equation}
and make and expansion of $a(w)$ and $r_0(w)$ of the form
\begin{eqnarray}
a(w)&=&a_2w^2+a_3w^3+\mathcal{O}(w^4),\label{eq:aw1n}\\
r_0(w)&=&r_H+r_1w+\mathcal{O}(w^2)\label{eq:r0nl},
\end{eqnarray}
with $a_2\neq 0$.
Therefore, by replacing $du=dw/\dot{a}(w)$ and \eqref{eqr0} in \eqref{eq:metricbondifr} and expanding in $w$ we obtain
\begin{equation}
    ds^2=Wdw^2+2dydw-r^2d\Omega^2,
\end{equation}
with $W$ given by
\begin{eqnarray}
   W=
\frac{f}{a(w)^2}
+ \frac{2}{a(w)} \left(\frac{d a(w)}{dw} y + \frac{d {r}_0}{dw}\right).
\end{eqnarray}
Taking into account the relations eqs.\eqref{eq:aw1n} and \eqref{eq:r0nl}, we obtain the following expansion of $W$ in powers of $w$:
}
{
\begin{widetext}
\begin{equation}\label{eq:extw}
    \begin{split}
    W=\frac{(a_2r_1)^2f''(r_H)+4r_1a^3_2}{w^2a^4_2}-\frac{-24a^5_2y+12a_3r_1a^3_2+(6a_3a^2_2r^2_1-6a^4_2r_1y)f''(r_H)-a^3_2r^3_1f'''(r_H)}{6w a^5_2}+\mathcal{O}(w^0),
    \end{split}
\end{equation}
\end{widetext}
with $f''(r_H)=\frac{2}{r_H^2}$, $f'''(r_H)=-\frac{12}{r^3_H}$.

In order the metric be regular at $w=0$,  the ${\cal O}(w^{-2})$ and ${\cal O}(w^{-1})$ contributions must vanish. The ${\cal O}(w^{-2})$ term is zero if $r_1=0$ or $r_1=-\frac{4a_2}{f''(r_H)}=-2r_H^2a_2$. However, the solution $r_1=0$ must be discarded because in that case the ${\cal O}(w^{-1})$ term of \eqref{eq:extw} cannot be made zero.  Taking knowledge of this, we solve for $a_3$ from the requirement of vanishing of the ${\cal O}(w^{-1})$ term, resulting in
$$a_3=-\frac{4}{3}\frac{f'''(r_H)}{(f''(r_H))^2}a^2_2=4r_Ha^2_2.$$

Hence, we have obtained that a family of regular coordinates at the neighborhood of $w=0$ is determined by
\begin{eqnarray}
a(w)&=&a_2(w^2+4r_Ha_2w^3)+\mathcal{O}(w^4),\label{eq:awfin}\\
r_0(w)&=&r_H-2a_2r^2_Hw+\mathcal{O}(w^2),\label{eq:r0fin}
\end{eqnarray}
with $a_2$ a free parameter. 
Note that the transformation given by \eqref{keywd} (with expansion \eqref{eq:wdoto}) and \eqref{q:rfinalwog} are compatible with the expansions \eqref{eq:awfin} and \eqref{eq:r0fin} by setting $a_2=\frac{E_0}{2r^2_H}$.
}
{Of course, there exist other possibilities that satisfy these relations. However, at we will see in Sec.\eqref{affnull}, these coordinates naturally appear in the affine-null metric formulation of the Einstein-Maxwell equations. An alternative proposal for the transformation $\{u,r\}\to \{w,y\}$ that can be checked to satisfy the relations \eqref{eq:awfin}-\eqref{eq:r0fin} and which is also global in the sense that admit maximal extension of the metric can be found in \cite{Kloesch1996}.}

{To finalize this Section, it is worthwhile to mention that that at the horizon $\mathcal{H}^+$, the Killing vector $\chi^a=(\partial_u)^a$ expressed in the $\{w,y\}$ basis coordinates reads  $\chi^a|_{\mathcal{H}^+}=-r_1(\partial_y)^a=2a_2r^2_H\ell^a$. In particular, $\chi^a$ never vanishes at $\mathcal{H}^+$, in agreement with the well known result that extremal black holes do not have  bifurcate Killing horizons\cite{WaldLR}. }
\section{Vaidya spacetimes}
Here we  present the outgoing and ingoing Vaidya solutions in the regular coordinates $(w,y)$. 

\subsection{The retarded Vaidya spacetimes}\label{vai_out}
 
Expressions for the outgoing Vaidya metric in regular $(w,y)$-type coordinates have already been discussed in the literature by Israel\cite{Israel1967} and Fayos et.al.\cite{Fayos}. However, our coordinate expression is not exactly the same and for completeness we present the form of this solution in the framework of Sec.\eqref{sec:reg_coord_setup}.

The Vaidya metric in Bondi coordinates  
is known as \cite{Vaidya1951}
\begin{equation}
ds^2=\left(1-\frac{2m(u)}{r}\right) du^2+2dudr-r^2 d\Omega^2
\end{equation}
In order to obtain the well behaved coordinates satisfying the requirements of Sec.~\ref{sec:intro}  we define \be w=-\exp(-k_Hu),\ee with $k_H=1/(4 m_0)$ where $m_0=\lim_{u\rightarrow\infty}m(u)$ and $r=\dot w y + 2m_0$ which gives
\be  
r(w,y)=2m_0-\frac{y w}{4m_0}.
\ee 
In the new coordinates, the metric becomes
\begin{equation}
\begin{split}
    ds^2 =& \Big(\frac{4 m_0}{w}\Big)^2\left(2-\frac{2m(w)}{r(w,y)}-\frac{r(w,y)}{2m_0}\right) dw^2\\
    &+2 dwdy-r^2(w,y) d\Omega^2,
    \end{split}
\end{equation}
and the Ricci tensor that follows from Einstein's equations is
\begin{equation}\label{eq:Rabs}
R_{ab}=\frac{ 8 m_0}{ wr(w,y)^2} \frac{dm}{dw}  {\ell}_a{\ell}_b=\rho_{\mathrm{out}}{\ell}_a{\ell}_b,
\end{equation}
with $\rho_{\mathrm{out}}$ representing the energy density of outgoing (scalar) radiation.
 Physically, its divergence would signal the presence of a firewall of outgoing radiation at the horizon detected as a physical divergence of $\rho_{\mathrm{out}}$ for any observers crossing the horizon. In principle there is not a constraint on the dependence of $m(w)$ with $w$, however, in our setting, as explained in Sec.\eqref{sec:reg_coord_setup}, we require regularity of the metric at the horizon. Such regularity therefore imposes that $m(w)= m_0+w^2 m_2+O(w^3)$. In terms of the Bondi mass $m(u)$ at $\sI^+$
 representing a fall-off of the form $\exp(-2k_H u)$ to the Schwarzschild geometry. 

The previous result has an intuitive meaning: in order for outgoing Vaidya radiation to scape to $\sI^+$ for late $u\to \infty$ it has to be sent with increasingly high local energy from the vicinity of the black hole horizon where the geometry imposes an ever increasing redshift. 
 
This effect has been pointed out before by Israel \cite{Israel1967} in terms of null coordinates that are very similar to the ones used here (See also Fayos et.al. where global extensions of this metric are discussed \cite{Fayos}). This effect is also reminiscent of the Christensen-Fulling  regularity conditions of the stress tensor in semiclassical studies of QFT on the Schwarzschild background \cite{ChristensenFulling} . 

\subsection{Ingoing Vaidya: collapsing null shell}\label{vai_in}
As a second application for Vaidya spacetimes  we consider a collapsing null shell. We assume that the spacetime  inside the shell is Minkowskian charted with the standard double null coordinates, the retarded time ${u_M=t_{\va M}-r_{\va M}}$, the advanced time ${v_M=t_{\va M}+r_{\va M}}$ and spherical angles $x^A_M = (\theta,\phi)$ so  that the metric is  
\begin{equation}
ds^2=dv_{\va M}du_{\va M}-r_{\va M}^2     q_{AB}dx^A_{\va M}dx^B_{\va M}.
\end{equation}
with \be2r_{\va M} = (v_{\va M}-u_{\va M})\label{eq:vum}.\ee
\begin{figure}[h!!!!!!!!!!!!!]
\includegraphics[height=7cm]{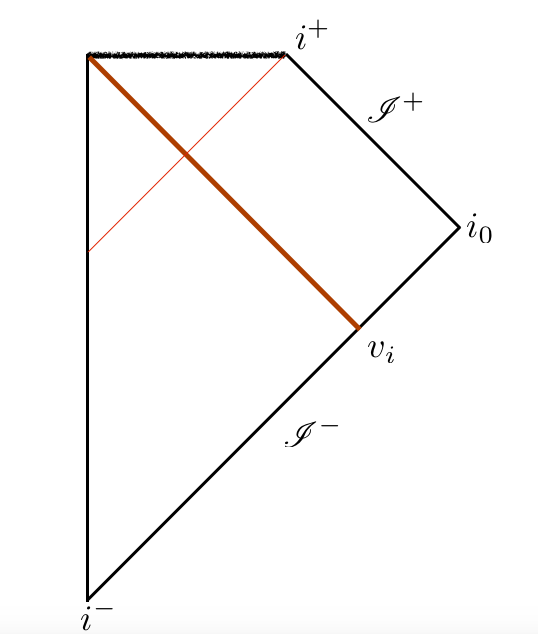}  
\caption{Penrose's diagram for the collapsing null shell. For values $v<v_i$ the spacetime is Minkowski, $v=v_i$ is the null shell and $v>v_i$ is the conformal diagram of a Schwarzschild  black hole.}
\label{fig:shell}
\end{figure}
The shell collapses (Fig.~\ref{fig:shell}) at the advanced time $v_{\va M} = v_i$ and its metric outside is described by a Schwarzschild metric in double null coordinates $(u, v, x^A)$
\begin{equation}
\begin{split}\label{eq:ds2vur}
ds^2=&(1-\frac{2m}{r})dvdu-r^2 q_{AB}dx^Adx^B 
\end{split}
\end{equation}
where $r$ is the areal radius function related to $u$ and $v$ by 
\be 2r^*=v-u,\label{eq:uvur}\ee 
with $r^*$ the tortoise coordinate 
\begin{equation}\label{eq:rtort}
r^* = r+ 2m\ln\left(\frac{r}{4m}-1\right).
\end{equation}

We require matching at the shell, i.e.: $x^A_{\va M} = x^A$, $v=v_M=v_i$ and the continuity of the areal radius function
\be{r(u,v_i)}= r_{\va M}(u_{\va M}, v_i) =\frac{v_i-u_{\va M}}{2} \label{eq:relacion}.\ee
From the transformation \eqref{eq:uvur}, the metric \eqref{eq:ds2vur} can be rewritten in terms of the Bondi coordinates (outgoing Eddington-Finkelstein coordinates) as
\begin{equation}
\begin{split}
    \label{vvaa}
ds^2
=&(1-\frac{2m}{r})du^2+2 dr du-r^2q_{AB}dx^Adx^B. 
\end{split}
\end{equation}
By replacing \eqref{eq:uvur}  and \eqref{eq:relacion} into \eqref{eq:rtort} (evaluated at $v=v_i$) we obtain
\begin{equation}
2r_* = (v_i-u)=(v_i-u_{\va M})+4m\ln\left(\frac{v_i-u_{\va M}}{4m}-1\right).
\end{equation}
Hence,
\begin{equation}\label{eq:uwum}
u=u_{\va M}-4m \ln\left(\frac{v_i-4m-u_{\va M}}{4m}\right).
\end{equation}
Now we define a new global coordinate
\begin{equation}\label{eq:intuw}
w\equiv u_{\va M}-v_i+4m,
    \end{equation}
where $w< 0$ outside the black holes horizon. In terms of $w$, \eqref{eq:uwum} reads   
\begin{equation}\label{eq:u(w)}
 {u=w+v_i-4m-4m \ln\left(-\frac{w}{4m}\right)}. 
\end{equation}
From this relation we find
\begin{equation}
\label{du_shell}
\begin{split}
    \frac{dw}{du} = \dot w=&\frac{1}{\left(1-\frac{4m}{w}\right)}
    =-\frac{k_H w}{1-k_Hw}\\
    =&-k_H{w}\left(1+k_H{w}+\left[k_Hw\right]^2+\cdots \right)    \end{split}
\end{equation}
 with $k_H=1/4m$, the surface gravity of the resulting Schwarzschild black hole. Note that \eqref{du_shell} is positive for  $w<0$.
 The  expansion in \eqref{du_shell} shows the regularity at $w=0$.
Applying the coordinate transformation \eqref{eq:u(w)} to \eqref{vvaa} while using  \eqref{du_shell}   yields
\begin{equation}\label{ds2_new_shell}
    \begin{split}
ds^2 =&(1-\frac{2m}{r})\left(1-\frac{4m}{w}\right)^2 dw^2+2 dr \left(1-\frac{4m}{w}\right) dw\\
&-r^2 d\Omega^2.
    \end{split}
\end{equation}
From eq.\eqref{eqq} and taking $r_0(w)=r_H=2m$, we obtain
\be\label{eessttee}
{r=\frac{y}{\left(1-\frac{4m}{w}\right)}+2m}
\ee
whose total differential is  
\be
dr=\frac{dy}{\left(1-\frac{4m}{w}\right)}-\frac{4m y dw}{w^2 \left(1-\frac{4m}{w}\right)^2}\;\;.
\ee
Note that from \eqref{eq:uwum}, $w$ and $u_M$ only differ in a constant, and therefore the associated affine parameters can also be chosen to agree, i.e. $r_M=y$. Moreover, using this identification between $r_M$ and $y$ and taking into account  eqs.\eqref{eq:relacion} and \eqref{eq:intuw}, we see that the null shell $v_M=v_i$ is described in the $\{w,y,x^A\}$ coordinates by $y=2m-\frac{w}{2}$.
This gives us 
the final form of the metric 
\begin{widetext}
\begin{equation}
ds^2 =\left\{\begin{array} {lll}\frac{-y(w^2-12mw+32m^2-8my)}{(4m-w)(yw+2mw-8m^2)}dw^2
+2 dy dw- \left(\frac{yw}{w- 4m }+2m\right)^2d\Omega^2\ \ \ \ \ y> \frac{4m-w}{2}\\
dw^2+2dw dy-y^2 d\Omega^2\ \ \ \ \ y\le \frac{4m-w}{2}
\end{array}\right.
\end{equation}
\end{widetext}
In the bottom we have the Minkowski metric where we can recognise $y=r_M$ is the affine parameter. 
The metric turns into the top one at the position of the shell. Notice that $r_M$ (the Minkowski radius) does not coincide with the affine parameter corresponding to the Bondi null hypersurfaces $u$=constant  that here we call $r$. 

\section{Einstein-Maxwell fields: A null affine characteristic formulation}\label{affnull}

In the previous sections, we have seen how to construct null regular coordinates starting from known spacetime solutions of the field equations written in Bondi type coordinates. In fact, the considered approach is valid regardless of the validity of Einstein's equations, as long as the requirements demanded in the introduction are met. A natural question is whether these coordinates can be obtained naturally by a direct solution of the field equations, without the need to previously go through another (Bondi) coordinate system. The answer to this question  is affirmative and leads towards the affine-null metric formulation of Einstein equations \cite{Win2013}. In particular, we present as an example, and for the first time in the literature, how to obtain the Reissner-Nordstr\"om solution directly at these regular coordinates starting with a characteristic formulation of Einstein's equations, i.e; giving certain data on a certain 2-sphere and on two null surfaces that intersect it orthogonally. 
\subsection{The characteristic initial value formulation}

\begin{figure}[h!!!!!!!!!!!!!]
\includegraphics[scale=0.4]{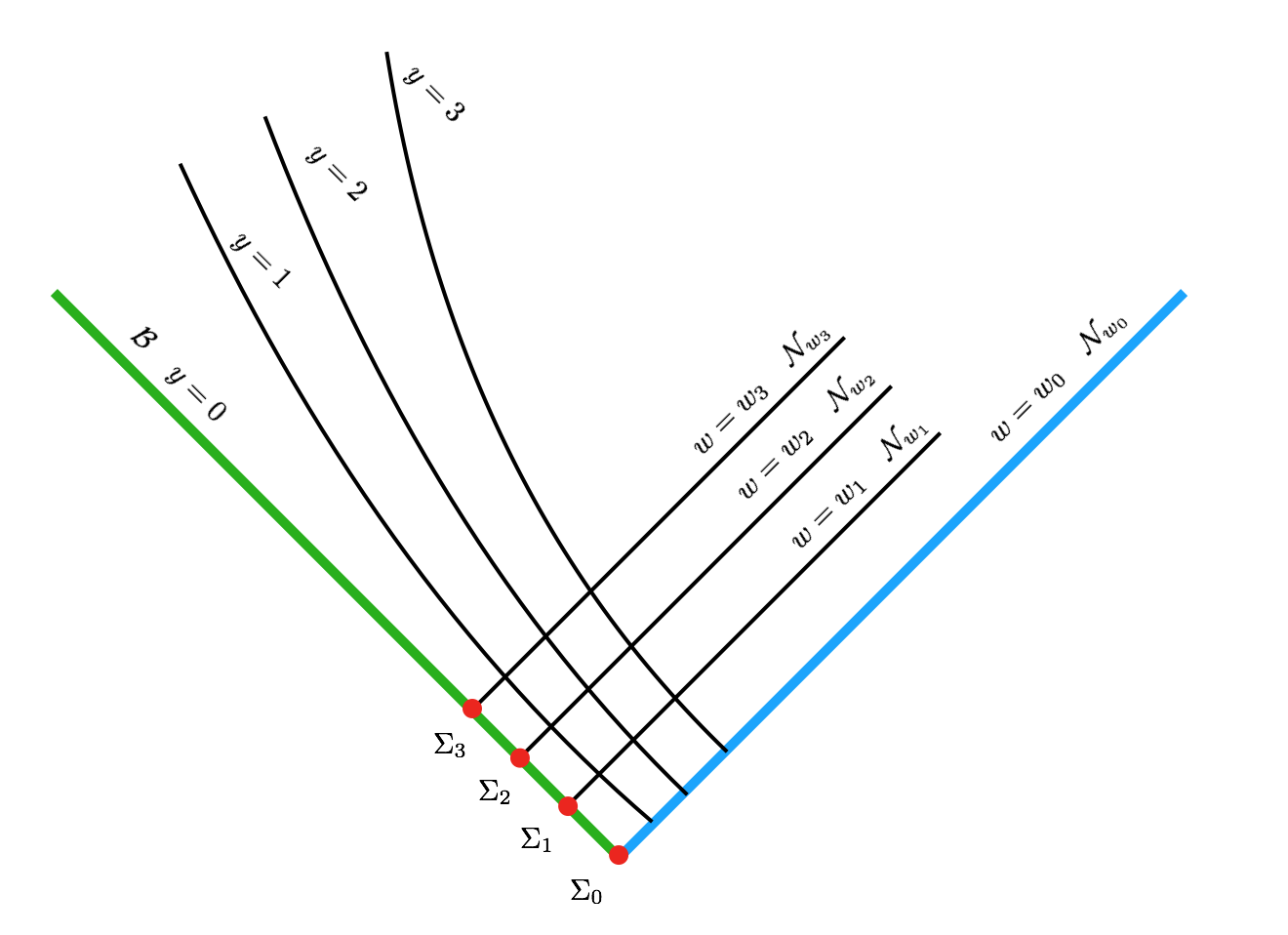} 
\caption{A graphical representation of the geometry for the spherically symmetric characteristic initial value formulation. The boundary surface $\mathcal{B}$ where $y=0$ is given in green while the initial data surface $\mathcal{N}_0$ where $w=w_0$ is lightblue. Every point in this diagram represents a two-dimensional sphere, in particular the red dots correspond to the two-dimensional common cross sections of the family of null hypersurfaces $\mathcal{N}_{w}$ with the boundary $\mathcal{B}$. In fact any other intersections of surfaces $y=const$ and $w=const$ are two-dimensional spaces. 
Note the hypersurfaces $y=const>0$ are in general no null hypersurfaces, like in a double null formulation where the surfaces $y=const$ would be displayed parallel to one another. }
\label{fig:afnull}
\end{figure}

In a spherically symmetric spacetime charted with coordinates $x^a = (w,y, x^A)$ consider a family of null hypersurfaces $\mathcal{N}_w = \{w=const\}$. The  surface forming rays of $\mathcal{N}_{w}$ are  parameterised with an affine parameter $x^1=y$. 
Suppose there is a designated null hypersurface $\mathcal{B}$ whose generators are orthogonal to those of $\mathcal{N}_{w}$.
At the common intersections $\Sigma_w$ of $\mathcal{N}_{w}$ and $\mathcal{B}$, we set $y=0$ (Fig.~\ref{fig:afnull}).
As any 4-dimensional spherically symmetric manifold can be represented as product of two dimensional spacetimes and round 2-spheres whose  total surface areas are given by $4\pi r^2$ using a radius function $r(x^a)$, we choose the area function $r$ such that the common intersections of $\mathcal{N}_w$ and $\mathcal{B}$ have the area $4\pi r^2(w, y=0)$ for every value of $w$.  The coordinates $x^A$ are the standard spherical angles of the spheres.

 A four dimensional metric with signature $-2$ adapted to the above is  given by  
\begin{equation}\label{eq:metrica}
     g_{ab}dx^adx^b=W dw^2 +2dwdy - r^2q_{AB}dx^Adx^B,
\end{equation}
and  inverse metric is 
\begin{equation}\label{eq:metricb}
    g^{ab}\partial_a\partial_b 
    = 2\partial_w\partial_y -W\partial_y^2-\frac{q^{AB}\partial_A\partial_B}{r^2},
\end{equation}
 where $q_{AB} = \mathrm{diag}(1, \sin^2\theta)$, $W|_{y=0}=0$ because  the coordinate surface $y=0$ is  the  null hypersurface $\mathcal{B}$. 
 We remark that the affine parameter $y$ has the gauge freedom $y\rightarrow A(w)+B(w)y$. On $\mathcal{B}$ we have the additional freedom to choose $w$ as an affine parameter along it generators giving the further condition $W_{,y}=0$\footnote{On $\mathcal{B}$, the null vector $n^a|_{\mathcal{B}}=(\partial_w)^a$ satisfy the geodesic equation $n^b\nabla_b n^a=-\frac{1}{2}W_{,y}n^a.$} \cite{Gomez2001, TM2019}. 
Ingoing ($n$) and outgoing ($\ell$) null vectors in terms of the metric fields are
 \begin{equation}
     \ell{^a} =(\partial_y)^a\;\;,\;\;
 n^a = (\partial_w)^a+\frac{W}{2}(\partial_y)^a,
 \end{equation}
 whose  expansion rates are\footnote{In order to preserve  notation of  previous works \cite{TM2019} (where ${y=\lambda}$), in this section we will do not make use of the GHP notation. However, note that in general $\Theta_\ell$ and $\Theta_n$ are related to the GHP scalars $\rho$, $\rho'$, $\epsilon$ and $\epsilon'$ by: $\Theta_\ell=-(\rho+\bar{\rho})+\epsilon+\bar{\epsilon}$, $\Theta_n=-(\rho'+\bar{\rho}')+\epsilon'+\bar{\epsilon}'$  respectively. In particular, in the case that $\ell$, $n$ are affine parametrized, $\epsilon+\bar\epsilon=\epsilon'+\bar\epsilon'=0$.} 
 \begin{eqnarray}
 \Theta_\ell &=&\nabla_a\ell^a= \partial_y \ln r^2,\\
 \Theta_n &=&\nabla_a n^a=\partial_w \ln r^2 + \frac{(Wr^2)_y}{2r^2}.
 \end{eqnarray}

 We consider the existence of a covector field  $A_a = (A_w, A_y, 0, 0)$ forming the Faraday tensor $F_{ab} = 2A_{[b,a]}$ of an electromagnetic field in vacuum.  The Faraday tensor has the gauge freedom in $A_a\rightarrow A_a + \chi_{,a}$ with the real scalar field $\chi$ allowing us to choose $A_y=0$ everywhere, by selecting a $\chi$ such that \cite{TW1966,Scholarpedia}
 \begin{equation}
     \chi(w,y) = -\int^y_0 A_{y}(w, \tilde y)d\tilde y.
 \end{equation}
 Hence, we have in the adapted null gauge
 \begin{equation}
     A_{a} = \alpha(w,y)dw.
 \end{equation}
 Here, we formulate a metric-based spherically symmetric characterisic initial value problem with respect to $\Sigma_w$, $\mathcal{B}$ and $\mathcal{N}_w$ for the Einstein-Maxwell equations for  the metric \eqref{eq:metrica}. This is an extension of \cite{Win2013,TM2019} to the Einstein-Maxwell case in spherical symmetry, while a formulation for an  Einstein-scalar field system can be found in \cite{Crespo2019}. 
 We provide initial boundary values for metric and Maxwell field on a null hypersurface $w_0= const$, on $\mathcal{B}$ given by $y=0$ and on the intersection  $\Sigma_{w_0}$ characterised by $w=w_0$ with $w_0$  and $y=0$. 
The field equations are $$R_{ab} = 8\pi (T_{ab}-\frac{1}{2}g_{ab}T^c_c)\;\;\;,\;\;\;
 \nabla_a T^{ab}=0,$$
 where $R_{ab}$ is the Ricci tensor and $T_{ab}$ is the energy momentum tensor (for negative metric signature) 
 $$T_{ab} = -\frac{1}{4\pi}\Big( F_{ac}F_b\;\!\!^c-\frac{1}{4}g_{ab}F^{cd}F_{cd}\Big)\;\;,\;\;T^c_c = 0,$$
 determined  by the Maxwell field $F_{ab}$.
 
 The divergence free condition of the energy momentum tensor gives the vacuum Maxwell equations $\nabla_a F^{ab} = 0$ which can be grouped into a  hypersurface equation
\begin{equation}
  0 = \frac{1}{r^2}(r^2\alpha_{,y})_{,y},  
\end{equation}
 assumed to hold everywhere on the family $\mathcal{N}_w$ and a
supplementary equation on $\mathcal{B}$
\begin{subequations}
\begin{equation}\label{eq:maxalphawB}
  0 = \frac{1}{r^2}(r^2\alpha_{,y})_{,w}\Big|_{\mathcal{B}}.  
\end{equation}
The supplementary equation holds everywhere provided the hypersurface equation are fulfilled everywhere \cite{TW1966}.

Similarily, the twice contracted Bianchi identities allow us to group the Einstein equations into one supplementary equation on $\mathcal{B}$, one  hypersurface equation with no $w-$derivatives and one evolution equation, respectively, for the metric fields  \cite{Win2013,TM2019}
 \begin{align}
\label{}
0    &   =\bigg( -\frac{2r_{,ww}}{r} \bigg)\bigg|_{\mathcal{B}}\label{eq:supp_r},\\
0    &   = -\frac{2r_{,yy}}{r}  ,\label{eq:hyp_r}\\
0    &   = [y + 2rr_{,w}-Wrr_{,y}]_{,y} - r^2\alpha_{,y}^2\label{eq:hyp_W}
\end{align}
\end{subequations}
Provided the hypersurface equation \eqref{eq:hyp_r} and evolution equation \eqref{eq:hyp_W} hold on the family $\mathcal{N}_w$, the supplementary equation \eqref{eq:supp_r} holds on $\mathcal{N}_w$, provided it is fulfilled on $\mathcal{B}$ \cite{TW1966}. 

The particular grouping of the Maxwell and Einstein equations  allows us to set up a characteristic initial boundary value problem for a family of null hypersurfaces $w>w_0=const$ with an initial null hypersurface $\mathcal {N}_{w=w_0}$, the null boundary surface $\mathcal{B}$ and the common intersection of $\Sigma_{w_0}$ of $\mathcal{N}_{w_0}$ and $\mathcal{B}$, (Fig. ~\ref{fig:afnull}).

 Since the spacetime  is spherically symmetric the shear tensors of two null hypersurfaces $\mathcal{B}$ and $\mathcal{N}_{w_0}$ must vanish and the intrinsic metrics on those null hypersurfaces must have the form $-r(w, y)^2q_{AB}$.
 Therefore, only the expansion rate of $\mathcal{N}_w$ and  $\mathcal{B}$ have nontrivial physical meaning. Since the two expansion rates evaluated on $\mathcal{B}$ and $\mathcal{N}_w$ determine the first derivatives of the conformal factor $r$ and as $r$ completely determines  the intrinsic properties of the two null hypersurfaces, it is natural to take $r$ and the  fields \footnote{The variable $N$ is called $\rho$ in \cite{Win2013,TM2019,Crespo2019}, we have chosen a different name here to not confuse it with the spin coefficients.}
 \begin{equation}\label{eq:var_rho_Thete}
N:=r_{,w}\;\;,\;\;
\Theta:=r_{,y},    
\end{equation}
as fundamental metric variables for a characteristic initial value formulation of the Einstein--Maxwell equations.
 Regarding the Maxwell field, only the derivatives  $\alpha_{,a}$ of the vector potential $A_a$ come into question as fundamental matter variables. Indeed, calculation of $F_{ab}$ reveals that only $\alpha_{,y}$ is needed to set up $F_{ab}$ in the chosen null gauge.

In fact, because $\Sigma_{w_0}$ is the common intersection of $\mathcal {B}$ and $\mathcal{N}_{w_0}$,
the fundamental matter and metric variables should be prescribed as independent constants on $\Sigma_{w_0}$,
\begin{equation}\label{eq:init_metric}
A:=\alpha_{,y}\Big|_{\Sigma_{w_0}}\;,\;\;
r_0:=r\Big|_{\Sigma_{w_0}}\;,\;\;
N_0=N\Big|_{\Sigma_{w_0}}\;,\;\;
\Theta_0=\Theta\Big|_{\Sigma_{w_0}}\;,\;\;
\end{equation}
Due to the spherical symmetry, it is possible to relate (before evaluation of the field equations on the boundary) $A$ with the charge $Q$ on $\Sigma_{w_0}$ in the following way;
The charge distribution on the  cross section $\Sigma_{w_0}$  is calculated according to 
\begin{equation}
\begin{split}
Q:=&Q|_{\Sigma_{w_0}} = \int_{\Sigma_{w_0}} j^a \Sigma_a 
= \frac{1}{4\pi}\int_{\Sigma_{w_0}} F^{ab}_{\;\;;b}d\Sigma_a\\
=&\frac{1}{8\pi}\int_{\Sigma_{w_0}} F^{ab}d\Sigma_{ab}, 
\end{split}
\end{equation}
where $Q$ is the total charge.
Using the surface element of the 2-surface $\Sigma_{w_0}$, $d\Sigma_{ab} = 2\ell_{[a}n_{b]}\sqrt{\det(g_{AB})}d\theta d\phi$, we find
\begin{equation}
    Q =\frac{1}{4\pi}\int_{\Sigma_{w_0}} F^{ab}r^2 \sin\theta d\theta d\phi = r_0^2A,
\end{equation}
Thus, we set
\begin{equation}
    \label{eq:init_Q}
  \alpha_{,y}|_{\Sigma_{w0}} = A = \frac{Q}{r^2_0},  
\end{equation}
hereafter.

Following \cite{Win2013,TM2019,Crespo2019}, introduction of
\begin{equation}\label{eq:Wdef}
   Y =  W -\frac{2 r_{,w}}{r_{,y}},
\end{equation}
casts \eqref{eq:hyp_W} into a hypersurface equation 
\begin{equation}
    0=1-(r\Theta Y)_{,y}-  r^2\alpha_{,y}^2.
\end{equation}

The hypersurface equation \eqref{eq:hyp_r} shows that its integration requires the knowledge of $r|_{\mathcal {B}}$ and $\Theta|_{\mathcal{B}}$, which are not known on $\mathcal{B}$ (the supplementary equations \eqref{eq:supp_r} only allows one propagate $N$ along $\mathcal{B}$ ). But if \eqref{eq:hyp_W} is evaluated on $\mathcal{B}$, we have 
\begin{equation} \label{eq:paramu}
0 = [1+2(r_{,w}r_{,y} +r r_{,wy}) - r^2 \alpha_{,y}^2]|_{\mathcal{B}},
\end{equation}
allowing us to propagate $\Theta$ along the boundary, i.e.
\begin{equation}\label{eqthetw} 
(2r \Theta)_{,w}|_{\mathcal{B}} = (-1   + r^2 \alpha_{,y}^2)|_{\mathcal{B}}.
\end{equation}
Moreover, \eqref{eq:paramu} allows us to find algebraically the mixed derivative
\begin{equation}
    \mu = r_{,wy}
\end{equation} 
everywhere on $\mathcal {B}$ provided we know $r$, $\Theta$, $N$ and $\alpha_{,y}$ on $\mathcal{B}$,
\begin{equation} \label{eq:muconst}
\mu|_{\mathcal{B}} =  \frac{-1-2N \Theta  + r^2 \alpha_{,y}^2}{2r}\Bigg|_{\mathcal{B}},
\end{equation}
In fact, evaluation of \eqref{eq:muconst} on the cross section $y=0$ and $w=w_0$ gives us
\begin{equation}
    \mu_0:=\mu \Big|_{\Sigma_{w_0}} = \frac{1 }{2r_{0}}\Big(\frac{Q^2}{r_0^2}  - 1 -2N_0\Theta_0\Big)\;\;.\label{eq:mu_def}
\end{equation}
where we used \eqref{eq:init_metric} and \eqref{eq:init_Q}.
Taking the $w$ derivative of eq.\eqref{eq:hyp_r}, we obtain an equation to propagate $\mu$ off the boundary $\mathcal{B}$
\begin{equation}
    \mu_{,y} = 0.
\end{equation}

We further note that the definition of $\mu = r_{,wy} = N_{,y}$ serves as an additional hypersurface equation to propagate $N$ off the boundary $\mathcal{B}$, 
\begin{equation}
    N_{,y} = \mu\;\;.
\end{equation}

Now we are in position to spell out the basic equations for the spherically symmetric Einstein-Maxwell system in a metric null-affine formulation.
Provided the data \eqref{eq:init_metric} and \eqref{eq:init_Q} on a cross section $\Sigma_{w_0}$, the boundary data are determined by the hierarchical set of equations 
\begin{subequations}\label{eq:bound_eq}
\begin{align}
    r_{,ww}\Big|_{\mathcal{B}}&=N_{,w}\Big|_{\mathcal{B}} = 0,\\
    (r^2\alpha_{,y})_{,w}\Big|_{\mathcal{B}} & = 0,\\
    (2r \Theta)_{,w}|_{\mathcal{B}} &= (-1  + r^2 \alpha^2_{,y})|_{\mathcal{B}},
\end{align}
\end{subequations}
together with the constraint \eqref{eq:muconst} and the initial value for $Y$,
\begin{equation}
    Y|_{\mathcal{B}} = -\frac{2N}{\Theta}\Big|_{\mathcal {B}}
\end{equation}
determined from \eqref{eq:Wdef} evaluated on $\mathcal {B}$.
The hypersurface equations are summarised by the following hierarchy
\begin{subequations}\label{eq:hyp_0ierarchy}
\begin{align}
    r_{,yy}&=\Theta_{,y}=0,\\
    (r^2\alpha_{,y})_{,y}&=0,\\
    (r\Theta Y)_{,y}&= 1- r^2\alpha_{,y}^2,\\
    \mu_{,y}&=0,\\
      {N_{,y}}&=\mu.
\end{align}
\end{subequations}
The missing metric field $W$ is found from the definition \eqref{eq:Wdef}.
\subsection{Solution of the hierarchy
of equations}
We now solve this system with initial values \eqref{eq:init_metric} and \eqref{eq:init_Q} and begin with the boundary equations \eqref{eq:bound_eq}. Its general solution  is
\begin{subequations}
\begin{align}
    r(w,0) &=r_0 + N_0(w-w_0),\label{eq:122a}\\
    N(w,0)& = N_0,\\
    \alpha_{,y}(w,0)&=\frac{Q}{r^2(w,0)}, \\
    \Theta(w,0)&=\frac{1}{r(w,0)}\Big[r_0\Theta_0 +\frac{1}{2}\Big(\frac{Q^2}{r_0 r(w,0)}-1\Big)(w-w_0) \Big],\label{eq:thetaw0}
\end{align}
together with the relations from the algebraic constraints
\begin{align}
    \mu(w,0) =&\frac{1}{2r(w,0)}\Big[-1-2N_0 \Theta(w,0)  + \frac{Q^2}{r^2(w,0)} \Big]\\
    Y(w, 0) =&-\frac{2 N_0 }{\Theta(w,0)}\label{eq_122f}
\end{align}
\end{subequations}

With these boundary values  at hand, the solution of the hypersurface equations  \eqref{eq:hyp_0ierarchy} is 
\begin{subequations}
\begin{align}
    \Theta(w,y)&=\Theta(w,0):=\Theta(w),\label{eq:thetawygen}\\
    r(w,y)& =r_0+N_0(w-w_0)
    +\Theta(w)y\\
    \alpha_{,y}(w,y)&=\frac{Q}{r^2(w,y)}\\
   Y(w,y)&=\frac{-2N_0 r(w,0)}{r(w,y)\Theta(w)}\nonumber\\ &+\frac{y}{r(w,y)\Theta(w)}\Big(1-\frac{Q^2}{r(w,0)r(w,y)}\Big)\Big]\\
   \mu(w,y)&=\mu(w,0):=\mu(w),\label{eq:sol_mu}\\
   N(w,y)& =\mu(w)y+N_0.\label{eq:rhoexp}
\end{align}
\end{subequations}

Having calculated $\Theta$, $Y$ and $N$ we are in position to find the missing metric field

\begin{equation}
\begin{split}\label{eq:WgeneralRN}
    W(w,y)=&\frac{y\Big[1+2\mu(w)r(w,y)+2N_0\Theta(w)\Big]}{r(w,y)\Theta(w)}\\&
    -\frac{yQ^2}{r^2(w,y)r(w,0)\Theta(w)},
    \end{split}
\end{equation}
which not only vanishes on $\mathcal{B}$, thus assuring that $\mathcal{B}$ is a null hypersurface, but also satisfies the other gauge condition  $W_{,y}|_\mathcal{B}=0$.

In spherical symmetry the mass of a system is given by the Misner-Sharp mass, $m$, defined covariantly via
\begin{equation}\label{eq:misner}
1-\frac{2m(r)}{r} = - g^{ab}r_{,a}r_{,b},
\end{equation}
where the minus sign in the right hand side is because of the  signature of spacetime. 
The Misner-Sharp mass is a special case for the quasi-local Hawking mass on a given 2-surface  $\Sigma$ \cite{Hawking1968}\footnote{The original reference gives a general form, \eqref{eq:HawkingEnergy} is its specialisation to maximal symmetry as e.g. found in \cite{Szabados2009}}
\begin{equation}\label{eq:HawkingEnergy}
    m_H(\Sigma)=\sqrt{\frac{A}{16\pi}}\left(1+\frac{1}{16\pi}\int_\Sigma \Theta_\ell\Theta_ndS\right),
\end{equation}
where $A$ is its area, $dS$ its surface area element and $\Theta_{\ell}$ and $\Theta_n$ are the expansion rates of two orthogonal null vectors that are orthogonal to $\Sigma$. In spherical symmetry the Hawking mass and Misner-Sharp mass coincide, which can be seen by evaluation of the two on $\Sigma_0$ resulting in
\begin{equation}\label{eq:msigma}
    m|_{\Sigma_0} = m_H(\Sigma_0) = \frac{r}{2}(1+g^{ab}r_{,a}r_{,b})|_{\Sigma_0} 
    =\frac{r_0}{2}(1+2N_0\Theta_0) 
\end{equation}
while using that 
\begin{equation}
\Theta_\ell\Big|_{\Sigma_0} = 2\frac{\Theta_0}{r_0}\;\;,\;\;
 \Theta_n\Big|_{\Sigma_0} =2\frac{N_0}{r_0}.
\end{equation}

 {From eqs.\eqref{eq:WgeneralRN}, \eqref{eq:misner} and \eqref{eq:msigma} we can also recover the well known relation between the quasilocal Misner-Sharp mass $m|_{\Sigma_0}$ and the Bondi mass $m_B$. More precisely, on an arbitrary  null  hypersurface $w=const,$ the Misner-Sharp mass \eqref{eq:misner} gives in the ``proper'' asymptotic limit the constant Bondi mass $m_B(w)=m_B=\text{const}$, 
\begin{equation}\label{eq:mBmqs}
\begin{split}
    m_B(w) :=& \lim_{y\rightarrow\infty\atop w= \text{const}}\frac{r}{2}(1+g^{ab}r_{,a}r_{,b})\\
    =& \frac{r_0}{2}(1+2N_0\Theta_0) +\frac{Q^2}{2r_0}=\\
    =& m|_{\Sigma_0}+\frac{Q^2}{2r_0}\equiv m_B=\text{const}.
    \end{split}
\end{equation}
From now on, we will refer to the Bondi mass $m_B$ simply as $m$.}

Until now the two surfaces $\mathcal{B}$ and  $\mathcal{N}_{w_0}$ have been arbitrary. With different choices for the corresponding initial data we can obtain the non extremal and extremal Reissner-Nordstr\"{o}m spacetimes. Let us consider these separate cases.
\subsubsection{The non-extremal Reissner-Nordstr\"{o}m metric}\label{subsub:nonext}
Let us choose now the data on $\Sigma_0$ such that $N_0=0$, and $\Theta_0\neq 0$. 
As follows of \eqref{eq:122a}, this choice implies that $r(w,0)=r_0=\text{const}\neq 0$.
Then, evaluation of \eqref{eq:sol_mu} shows the constancy of the field $\mu$ everywhere
\begin{equation}\label{eq:mu0}
    \mu_0 = \mu(w)= - \frac{r^2_0-Q^2}{2r^3_0}.
\end{equation} 
Consequently, the null hypersurface $\mathcal{B}$  is free of  expansion, i.e. $\Theta_{n}|_{\mathcal{B}} = 0$.

With respect to this  expansion-free null boundary, we see that the metric functions reduce to
\begin{eqnarray}
r&=&r_0+y\left[\Theta_0+\mu_0(w-w_0)\right],\\
W&=&\frac{2y(\mu^2_0(w-w_0)+\mu_0\Theta_0)+4r_0\mu_0+1}{\{r_0+[\Theta_0+\mu_0(w-w_0)]y\}^2}y^2 .
\end{eqnarray}
Also note that from \eqref{eq:thetawygen} and \eqref{eq:thetaw0} follows the existence of  a null hypersurface $w = \hat w$ in $\mathcal{N}_{w}$ such that  $\Theta(\hat w,y)=0$, meaning 
\begin{equation}
    \Theta(\hat w,y) =\frac{1}{r_0}\Big[r_0\Theta_0 +\frac{1}{2}\Big(-1+\frac{Q^2}{r_0^2}\Big)(\hat w-w_0) \Big]\stackrel{!}{=}0,
\end{equation}
which implies 
\begin{equation}
\hat w=\frac{-\Theta_0+\mu_0 w_0}{\mu_0}.    
\end{equation}
The coordinate transformation  $w\rightarrow \tilde{w}= w-\hat w$ shifts   the affine parameter $w$ of the geodesics on $\mathcal{B}$ such that
 the metric components have a particular simple form
\begin{eqnarray}
r&=&r_0+\mu_0 wy,\\
W&=&\frac{1+2\mu_0(r+r_0)}{r^2}y^2 ,
\end{eqnarray}
independent of the  values of $\Theta_0$ and $w_0$ where 
we have dropped the tilde  in order do not overcomplicate  the notation.

Using eq.\eqref{eq:mBmqs} we can relate the constant $r_0$ with the Bondi mass $m$ of the metric: 
\begin{equation}
\begin{split}
   \frac{Q^2+r^2_0}{2r_0}= m.
    \end{split}
\end{equation}
with solutions 
\begin{equation}
    r_0(\epsilon)=m+\epsilon\sqrt{m^2-Q^2},
\end{equation}
where $\epsilon=\pm 1$.  We  see that $r_{0}(\epsilon=1)$ is nothing else then the value(s) $r^+_H$ of the event horizon of the Reissner-Nordstr\"om solution. Note also, that the value of $\mu_0$ given by \eqref{eq:mu0} agrees with the value  of minus the surface gravity, $-k_H$, of this spacetime. Indeed, this solution is exactly the same as  \eqref{eq:rgene} presented in the Sec.\eqref{non_ex} which was obtained using the coordinate transformation from Bondi to the regular ones $\{w,y\}$.

\subsubsection{The extremal Reissner-Nordstr\"{o}m metric}
{For the extremal case $Q=m$ we
can proceed in two different ways. 
\\

\underline{Data I}:
{Choosing again $N_0=0$ and $\Theta_0\neq 0$ on $\Sigma_0$, we can use the results of the previous subsection \eqref{subsub:nonext} obtaining for the extremal case:
\begin{eqnarray}
 \mu(w)&=&\mu_0=0,\\
 r_0&=&m,\label{eq:r0b}\\
 r&=&m+\Theta_0y,\\
 W&=&\frac{y^2}{r^2},\\
 \Theta(w,y)&=&\Theta_0.\label{eq:Theta0b}
\end{eqnarray}
}
{We recognize that the resulting coordinates $(w,y)$ are basically the Bondi coordinates $(u,r)$ after a rescaling of $u$ and a reparametrization of the affine parameter $r$: $w=\Theta_0u$, $r=m+\Theta_0y$, with metric \begin{equation}
    ds^2=\left(1-\frac{m}{r}\right)^2du^2+2dudr-r^2d\Omega^2.
\end{equation}
However, we know that this coordinate system is not regular at the horizon $\mathcal{H}^+$, in particular there is not a $w=\hat{w}$ value where $\Theta(\tilde{w},y)=0$. Hence, in order to find a regular coordinate system at $\mathcal{H}^+$ we must to set the initial data diferently.

Let us also remark that in the limit $\Theta_0\to 0$ on $\Sigma_0$ we obtain from eqs.\eqref{eq:r0b}-\eqref{eq:Theta0b},
\begin{eqnarray}
 r&=&m=Q,\\
 W&=&\frac{y^2}{m^2}=\frac{y^2}{Q^2},\\
 \alpha&=&\frac{y}{Q},
\end{eqnarray}
giving as a limit the metric 
\begin{equation}\label{eq:148}
    ds^2=\frac{y^2}{Q^2}dw^2+2dwdy-Q^2d\Omega^2,
\end{equation}
with electromagnetic potential vector $A_a=\frac{y}{Q}dw$.

This metric is the well-known Bertotti-Robinson solution which represent a shear and expansion free conformally flat solution of the Einstein-Maxwell equations with uniform electromagnetic field\cite{Robinson,Bertotti,Dolan,Tariq}.
\\

\underline{Data II}:
Let us choose both $N_0$ and $\Theta_0$ different from zero on $\Sigma_0$. From eq.\eqref{eq:mBmqs} we obtain for the extremal case
\begin{equation}
    \Theta_0=-\frac{(m-r_0)^2}{2N_0r_0^2}.
\end{equation}
After replacing this expression into the eqs.\eqref{eq:122a}, \eqref{eq:thetaw0} and \eqref{eq:thetawygen} we obtain
\begin{equation}
    \Theta(w,y)=\Theta(w,0)=-\frac{[N_0(w-w_0)-m+r_0]^2}{2N_0[r_0+N_0(w-w_0)]^2}.
\end{equation}
Therefore, there exist  a null hypersurface $w = \hat w$ in $\mathcal{N}_{w}$ such that  $\Theta(\hat w,y)=0$, which implies
\begin{equation}
\hat{w}=\frac{m-r_0+N_0w_0}{N_0}.
\end{equation}
As in \eqref{subsub:nonext}, the coordinate transformation  $w\rightarrow \tilde{w}= w-\hat w$ shifts the affine parameter $w$ of the geodesics on $\mathcal{B}$ such that from the set of equations \eqref{eq:122a}-\eqref{eq_122f}, \eqref{eq:thetawygen}-\eqref{eq:rhoexp} and \eqref{eq:WgeneralRN} we obtain for the metric components $r$ and $W$ the following expressions (dropping again the tilde  in $w$ for simplicity in the notation): 
\begin{widetext}
\begin{eqnarray}
r&=&m+N_0w-\frac{N_0w^2y}{2(N_0w+m)^2},\label{eqrrn}\\
W&=&-\frac{4(2w^4N_0^4+5w^3N_0^3m+3w^2N_0^2m^2-wN_0m^3-w^3yN_0^2-m^4)my^2}{(N_0w+m)(2w^3N_0^3+6w^2N_0^2m+6wN_0m^2+2m^3-w^2N_0y)^2},\label{eqwrn}
\end{eqnarray}
\end{widetext}
which agrees with the solution presented in Sec.\eqref{ex} after the identification  $N_0=-E_0$\footnote{This identification follows from the definition of $N$ in eq.\eqref{eq:var_rho_Thete} and eq.\eqref{eq:rel} (with $w\equiv \lambda$); both valued at $\Sigma_0$.}. 
Note again, that near the horizon $\mathcal{H}^+$, $r$ and $W$ behave as:
\begin{eqnarray}
    r&=&m+\mathcal{O}(w),\\
    W&=&\frac{y^2}{m^2}+\mathcal{O}(w),
\end{eqnarray}
in correspondence with the well-known result that an extremal Reissner-Nordström black hole near the horizon looks like the Bertotti-Robinson metric eq.\eqref{eq:148}\cite{Carter2}.

\section{Summary and outlook}
In this work, we have discussed two different approaches to study coordinate charts of black hole spacetimes that are regular at the black hole horizon and at large distances towards null infinity. In particular, we discuss these charts in spherical symmetry. The first approach consists in defining a suitable   coordinate pair $(w,y)$, which obeys geometric conditions as specified  in Sec.~\ref{sec:reg_coord_setup}. Of particular importance in the construction of this pair  is the Bondi coordinate pair $(u, r)$  where $u$ takes the value of the retarded time in the limit towards null infinity and $r$ is the affine parameter on surfaces $u=const$. The pair $(w,y)$ is then constructed by remapping the affine parameter $r$ of  the surfaces $u=const$ to another  parameter $y$. 
Unlike $(u,r)$,  $(w,y)$ can be defined from the horizon to null infinity when caustics are not present.
Within this formalism, we find i) 
regular representations of static and spherically symmetric black holes, including solutions for extremal/non-extremal Reissner Nordstr\"{o}m black holes, ii) the outgoing Vaidya metric depending  on a mass function $m(w)$ that has an extremum at $w=0$ and iii) an ingoing Vaidya solution of a collapsing shell. All those  solutions are explicit with respect to the $(w,y)$ pair, allowing thus to write exemplary black hole spacetimes using null coordinates in an explicit, rather than implicit, way  from the horizon to null infinity.

In the second approach, we have started out with a general  affine-null coordinate system. In this affine-null chart, we set up a characteristic initial value problem\footnote{A past value problem is set up in a similar way.} for  the field equations in spherical symmetry for an Einstein-Maxwell system. The derived equations form a hierarchical system of equations that can be solved with  data given on a common intersection of two null hypersurfaces. There are in principle four free parameters that can be specified in  the spherically symmetric case discussed here. Nevertheless, after fixing the values of the expansion rates of the in- and outgoing  null vectors at the common intersection,  the final solution provides us with the Reissner-Nordstr\"{o}m black hole 
whose line element is given by \eqref{eq:metricsphegen} for the non-extremal case, and determined by the metric components eqs.\eqref{eqrrn} and \eqref{eqwrn} in the extremal situation. 

For the non extremal case, we can pursue with a compactification procedure for \eqref{eq:rgene} like for \eqref{BH_israel}. If we make the rescaling $w\rightarrow k_H w$ and $y\rightarrow (k_H\Upsilon)^{-1}$ with $k_H$ 
the surface gravity of Reissner-Nordstr\"{o}m black hole together with a suitable conformal factor $\Omega  = k_H\Upsilon$, the resulting conformal metric has the expansion at $\mathscr{I}^+$, i.e. at $\Upsilon=0$, like \eqref{metricOffScri}. 

 To our knowledge, our presentation  is the first to construct a Reissner-Nordstr\"{o}m black hole in Israel like coordinates directly by solving the Einstein equations using a characteristic initial value  problem, while other approaches  relied on coordinate transformations \cite{Israel1966} or using two-dimensional generalized dilaton gravity models \cite{Kloesch1996}(also see for discussion \cite{Blau}).

Both of the two approaches have their regime in which they are most useful. The framework starting out with a  Bondi coordinate $u$ is useful (as we have demonstrated) if particular solutions of the Einstein equations are known. Then, the null coordinate $w$ and its  affine parameter $y$ can be directly constructed locally. 
As we have restricted the framework to spherical symmetry,  further work is needed to extend it to more general spacetimes, for example axisymmetric spacetimes.
Regarding the coordinates, we expect a condition not only linking $r$ with $y$, but also also additional relationships  between the angular coordinates  parameterising the cuts $u=const$ at different values of $r$ with those of cuts of $w$ for given values of $y$ \cite {doubleKerr}. Indeed in their discussion of the gravitational wave memory effect of boosted Kerr-Schild black holes \cite{2018CQGra..35c5009M,2019CQGra..36i5009M}, the authors  have shown  in \cite{2019CQGra..36i5009M} using a  Penrose compactification scheme on uncharged Kerr-Schild metrics, that an asymptotic Bondi frame can only be properly constructed if the angular coordinates issue is taken into account. 
It remains to be of further study, how the condition of Sec.~\ref{sec:reg_coord_setup} need to be  specified for more general systems. 

The second approach is most useful numerically, i.e. when we wish to find a solution of the Einstein equation given certain initial values. Recently, Crespo and collaborators \cite{Crespo2019} have used an affine-null  formulation in spherical symmetry and with a scalar field to study the Choptuik critical solution. 
The authors have integrated the Einstein-scalar field equations numerically, also using a hierachical set of equations. 
Their equations, however, are different from ours, because the matter terms enter into the $r-$hypersurface equation \eqref{eq:hyp_r} rather than the $W-$hypersurface equation \eqref{eq:hyp_W} as in our case. 
Deriving a hierarchy is thus different and it remains to be seen how more general affine-null systems can be treated.

\subsubsection*{Acknowledgements}

We acknowledge financial support from CONICET, SeCyT-UNC, Foncyt and by the Agence Nationale de la Recherche; grant
ANR-06-BLAN-0050. A.P. was supported by {\em l'Institut Universitaire de France}. T. M. is supported by FONDECYT de iniciaci\'on 2019 (Project No. 11190854) of the Chilean National Agency for the Science and Technology (CONICYT).

\begingroup\raggedright\endgroup

\end{document}